\documentclass[9pt]{article}
\usepackage{spconf}
\usepackage{amsmath,graphicx,amssymb}
\usepackage{caption}
\usepackage{subcaption}
\usepackage{url}
\usepackage{multirow}
\usepackage{algorithm}
\usepackage{algpseudocode}
\usepackage{dblfloatfix}

\newcommand{\myfigwidth}{0.285\textwidth}

\title{On Class Separability Pitfalls In Audio-Text Contrastive Zero-Shot Learning}
\name{Anonymous\thanks{Anonymous.}}
\address{Anonymous}

\name{
Tiago Fernandes Tavares$^{\sharp}
$\thanks{Authors thank the Insper-UIUC Research Partnership for funding.}
\qquad
Fabio Ayres$^{\sharp}$
\qquad
Zhepei Wang$^{\flat}$
\qquad
Paris Smaragdis$^{\flat}$
}
  
\address{
$^{\sharp}$INSPER Institute of Teaching and Research \\
$^{\flat}$University of Illinois at Urbana-Champaign
}

\usepackage{tikz}
\usetikzlibrary{positioning, calc}
\usetikzlibrary{shapes.geometric, arrows}
\tikzstyle{start} = [rectangle, rounded corners, minimum width=2cm, minimum height=1cm,text centered, draw=black, fill=gray!30]
\tikzstyle{startstop} = [rectangle, rounded corners, minimum width=2cm, minimum height=1cm,text centered, draw=black, fill=gray!30]
\tikzstyle{invisible} = [rectangle, text centered, draw=none]
\tikzstyle{process} = [rectangle, minimum width=1cm, minimum height=1cm, text centered, draw=black, fill=white]
\tikzstyle{component} = [rectangle, rounded corners, minimum width=1cm, minimum height=0.5cm, text centered, draw=black, fill=yellow!30]
\tikzstyle{decision} = [diamond, minimum width=2cm, minimum height=0.5cm, text centered, draw=black, fill=green!30]
\tikzstyle{io} = [trapezium, trapezium left angle=70, trapezium right angle=110, minimum width=3cm, minimum height=1cm, text centered, draw=black, fill=blue!30]
\tikzstyle{arrow} = [thick,->,>=stealth]
\tikzstyle{dasharrow} = [dashed,->,>=stealth]
\tikzstyle{state} = [circle, minimum width=1cm, minimum height=1cm, text centered, draw=black, fill=blue!30]

\begin{document}

\ninept
\maketitle
\begin{abstract}
Recent advances in audio-text cross-modal contrastive learning have shown its potential towards zero-shot learning. One possibility for this is by projecting item embeddings from pre-trained backbone neural networks into a cross-modal space in which item similarity can be calculated in either domain. This process relies on a strong unimodal pre-training of the backbone networks, and on a data-intensive training task for the projectors. These two processes can be biased by unintentional data leakage, which can arise from using supervised learning in pre-training or from inadvertently training the cross-modal projection using labels from the zero-shot learning evaluation. In this study, we show that a significant part of the measured zero-shot learning accuracy is due to strengths inherited from the audio and text backbones, that is, they are not learned in the cross-modal domain and are not transferred from one modality to another.
\end{abstract}

\section{Introduction}
\label{sec:intro}

Zero-shot learning is a machine-learning paradigm in which classes in the training and testing set do not overlap~\cite{pourpanah_review_2022}, effectively requiring machines to learn using the similarity of class-related attributes~\cite{lampert_learning_2009}. 
This methodology allows the identification of classes that are not present in the training set by associating them with attributes learned from the classes in the training set.

More recent systems for cross-modal zero-shot learning rely on building a cross-modal (or mixed-domain, or latent space) embedding that semantically represents items in either of the initial modalities~\cite{watanabe_query-by-blending_2019,xian_zero-shot_2020,radford_clip_2021,elizalde2022clap}. For example, in an audio-text cross-modal embedding, the sound of a car should be close to the sound of a truck, and they are respectively close to the representations of the words ``car'' and ``truck''. Thus, a similar item, like the sound of a motorcycle, would be represented close to the cross-modal embedding of the word ``motorcycle'', hence allowing retrieval of classes not present in the audio training set.

Cross-modal zero-shot learning depends on an initial unimodal representation in which items are semantically organized. This can be obtained by using word embeddings~\cite{xie_zero-shot_2019,watanabe_query-by-blending_2019} or language models~\cite{radford_clip_2021,elizalde2022clap} for text representation, and feature extractors~\cite{watanabe_query-by-blending_2019} or pre-trained~\cite{xie_zero-shot_2019,radford_clip_2021,elizalde2022clap,deshmukh2022audio} neural networks for image or audio. The systems that generate unimodal representations are called \textit{backbones}, and the neural networks that map the backbone outputs to the cross-modal space are called \textit{projectors}.

In the cross-modal space, embeddings related to the media (audio or image) and the text data for any media-text pair should be similar. Hence, the cross-modal space can be understood as a translation layer between the modes. In these conditions, these representations allow item comparisons by vector similarity in the cross-modal domain, thus allowing the extrapolation of their training data to similar, yet unseen, concepts~\cite{radford_clip_2021,elizalde2022clap,Yusong2023-laionclap}.

Training backbone networks for unimodal representations, and fitting projectors to generate cross-modal embeddings, are data-intensive tasks. They commonly rely on pre-trained backbones and on obtaining large datasets from scraping. These strategies are important, but can lead to inadvertently contaminating the labels in the training set with labels present on the testing set~\cite{choi_zero-shot_2019,xian_zero-shot_2020,mercea_audiovisual_2022}. Consequently, the resulting machines are not really tested in a zero-shot setting~\cite{choi_zero-shot_2019}; rather, they become powerful classifiers towards a large amount of possible known labels.

The data overlap problem is well known. It has been avoided in prior work~\cite{xie_zero-shot_2019} and its effects have been evaluated in terms of classification accuracy~\cite{choi_zero-shot_2019}. However, more recent work has not fully addressed the data overlap aspect~\cite{elizalde2022clap}, or used pre-trained networks that have seen labels in a supervised manner~\cite{wu2023}. Despite their importance, the exact effects of data leakage - other than their impact on classification accuracy - still require further discussion.

In this paper, we explore a methodology for assessing the impact of data leakage in cross-modal contrastive learning. We investigate the relationship between the zero-shot classification accuracy and the structure of the embedding spaces by evaluating projections using T-SNE visualizations~\cite{tsne}, silhouette scores~\cite{Rousseeuw1987} and an neighborhood-based measure for topological similarity. Our study reveals mechanisms by which data leakage affects embedding spaces and leads to undue higher accuracy. 
We use the Contrastive Language-Audio Pretraining (CLAP)~\cite{elizalde2022clap} model as a case study, demonstrating that the process discussed in this paper is capable of identifying inadvertent data leakage and the effects of cross-modal training in the performance of zero-shot classification. 

The present work contributes to the current understanding of cross-modal contrastive zero-shot learning with the following insights: the zero-shot classification accuracy is closely linked to the separability of classes in the audio side of the cross-modal domain; cross-modal embeddings derived from text and audio may not necessarily exhibit similar topological structures; and projectors may struggle to differentiate classes that are poorly separated in one modality, even if they are clearly separated in the other modality.

\section{Zero-shot learning model}
\label{sec:model}

We base this work on a reproduction of the CLAP audio-text contrastive learning framework~\cite{elizalde2022clap}. This topology has been applied in further work in zero-shot classification with some modifications~\cite{deshmukh2022audio}, and as conditioning for generative neural networks~\cite{ramesh_hierarchical_2022}. In this section, we review the CLAP topology and its implementation details, and highlight possible sources of data leakage in its training process.

\subsection{Contrastive learning}
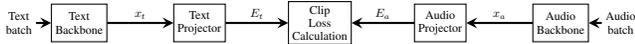
\begin{figure}[h]
\centering
\begin{tikzpicture}[node distance=1.6cm,scale=0.4, every node/.style={scale=0.5,align=center}]

\node (inputT) [invisible] {Text\\batch};
\node (textNetwork) [process, right of=inputT] {Text\\Backbone};
\node (textEmbeddings) [invisible, right of=textNetwork] {};
\node (textProjector) [process, right of=textEmbeddings] {Text\\ Projector};
\node (textProjections) [invisible, right of=textProjector] {};

\node (C2) [process, right of=textProjections] {Clip\\Loss\\ Calculation};

\node (audioProjections) [invisible, right of=C2] {};
\node (audioProjector) [process, right of=audioProjections] {Audio\\Projector};
\node (audioEmbeddings) [invisible, right of=audioProjector]{};
\node (audioNetwork) [process, right of=audioEmbeddings] {Audio\\Backbone};

\node (inputA) [invisible, right of=audioNetwork] {Audio\\batch};


\draw [arrow] (inputT) -- (textNetwork);
\draw [arrow] (textNetwork) -- (textProjector) node[midway,above]{$x_t$};
\draw [arrow] (textProjector) -- (C2) node[midway,above]{$E_t$};

\draw [arrow] (inputA) -- (audioNetwork);
\draw [arrow] (audioNetwork) -- (audioProjector) node[midway, above]{$x_a$};
\draw [arrow] (audioProjector) -- (C2) node[midway,above]{$E_a$};


\end{tikzpicture}
\caption{Contrastive learning topology. Each unimodal branch is made of a pre-trained neural network (NN) backbone (CNN14 yielding $x_a$ for audio, and BERT yielding $x_t$ for text), and projectors are MLP networks yielding $E_a$ and $E_t$ respectively for audio and text. }
\label{fig:topology}
\end{figure}

As shown in Figure~\ref{fig:topology}, the CLAP topology consists of two unimodal branches, each with a backbone pre-trained neural network (CNN14~\cite{cnn14} for audio, BERT~\cite{bert} for text) followed by an MLP projector. The backbone networks yield unimodal embeddings for text ($x_t$) and audio ($x_a$), which are further projected by MLP networks into cross-domain embeddings $E_t$ and $E_a$. The cross-domain embeddings are normalized to unit $\ell_2$ norm. The weights of projectors and neural networks are optimized as to minimize the CLIP loss~\cite{radford_clip_2021,elizalde2022clap}.


We have found this topology to be very sensitive to parameter variations in the training process. Because of that, after obtaining similar results to those in the original work~\cite{elizalde2022clap}, we used the following strategies to avoid ill-fitting: (a) Text and audio backbones were initialized using pre-trained weights respectively from HuggingFace\footnote{\url{https://huggingface.co/}, model \textsc{bert-base-uncased}} and PANNs\footnote{\url{https://github.com/qiuqiangkong/audioset_tagging_cnn} model \textsc{Cnn14\_mAP=0.431.pth}}. (b) Started training with frozen weights in the backbone for 5~epochs. (c) While frozen, use Adam optimizer with learning rate~$10^{-4}$. (d) Unfreeze both backbone networks and monitor loss in the validation set after epoch~5. (e) After unfreeze, use Adam optimizer with learning rate~$10^{-5}$ until the loss in the validation set stops decreasing. (f) During all training, limit gradient magnitude to~$100$. This avoids non-finite results while applying backpropagation. (g) During all training, limit values in~$C$ to the range $[-100,100]$. This constrains the impact of the temperature factor~$\tau$. (h) Gradient decreases on a plateau by a factor of~$10^{-1}$ with patience 10 according to the loss on the validation set. (i) Audio material used for training was extracted from a random segment of original files, resampled to $44100$~Hz (following the original implementation \cite{elizalde2022clap}), clipped (or zero-padded, if necessary) to $5$~s, and normalized to zero mean and unit variance. (j) The batch size $b$ was fixed at 128.

\subsection{Datasets}
The dataset used for training CLAP consists of the union of four audio-text datasets: Clotho~\cite{clotho}, Audiocaps~\cite{audiocaps}, FSD50K \cite{fonseca2022FSD50K}, and MACS~\cite{macs}. Their audio files were paired with each of the corresponding captions (that is, an audio file can appear more than once in the dataset, each time being paired with a different caption). We randomly split 20\% of the audio-text pairs for validation.

Test data consists of the ESC50 dataset~\cite{piczak2015dataset}. It contains sounds labeled into 50 classes. It was used in a zero-shot learning evaluation using the similarity between their audio embeddings and their label (text) embeddings~\cite{elizalde2022clap}.


Furthermore, we observed that labels in the test data were present both in the pre-training dataset~\cite{audioset} and the training data. To avoid leakage, we removed all audio examples whose captions contain words that were present in the ESC50 class set. Also, we experimented with using a random initialization of CNN14 instead of the pre-trained weights.



The original datasets are labeled \textit{dirty}, and their filtered versions are labeled \textit{clean}, in reference to the existence of data leakage through the captions. In total, there are two possibilities for training (``dirty'' and ``clean''), and three possibilities for pre-training (``dirty'', ``clean'', and ``none'' for random initialization of weights). This results in six different experimental setups. Hence, the variation ``dirty/dirty'' corresponds to the exact recipe described in \cite{elizalde2022clap}, whereas ``clean/clean'' corresponds to pre-training and training with filtered datasets.


\section{Experiments and results}
\label{sec:experiments}

The experiments conducted in this paper consist in assessing the relationships between the zero-shot classification accuracy and specific characteristics of the embeddings generated by the trained CLAP models. The accuracy in the ESC50 zero-shot learning task is reported in Table \ref{tab:results_esc50}, showing that, as more sources of data leakage are removed, the zero-shot accuracy clearly decreases. The performance of ``dirty/dirty'' is close to the one reported in the original paper \cite{elizalde2022clap}, whereas the ``none/clean'' presents accuracy closer to the ones reported in prior research in zero-shot learning for audio \cite{choi_zero-shot_2019,xie_zero-shot_2021}. Importantly, ``clean/clean'' outperforms work preceding CLAP \cite{elizalde2022clap}, but by a smaller margin than ``dirty/dirty''.



\begin{figure*}[t]
\centering
\begin{subfigure}[b]{\myfigwidth}
\centering
\includegraphics[width=\textwidth]{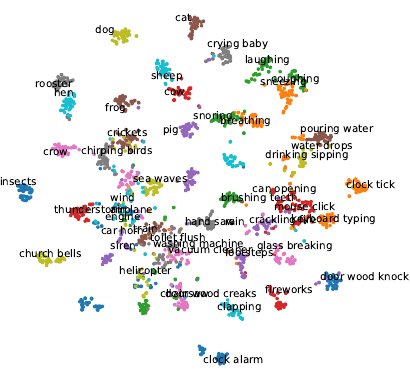}
\caption{Dirty training, dirty pre-training.}
\label{fig:ea_dirty_dirty}
\end{subfigure}
\begin{subfigure}[b]{\myfigwidth}
\centering
\includegraphics[width=\textwidth]{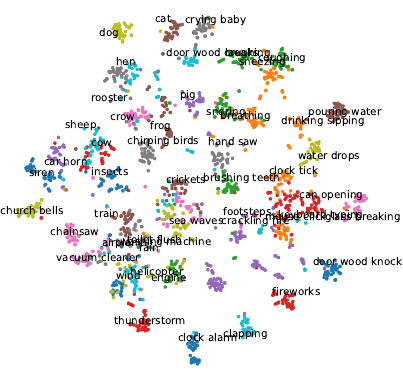}
\caption{Dirty training, clean pre-training.}
\label{fig:ea_dirty_clean}
\end{subfigure}
\begin{subfigure}[b]{\myfigwidth}
\centering
\includegraphics[width=\textwidth]{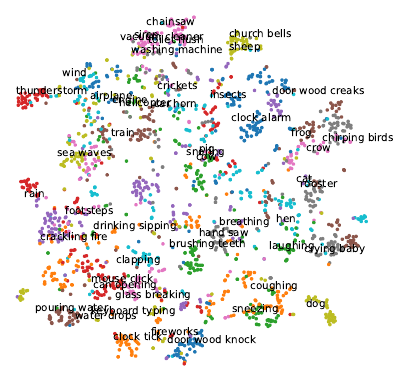}
\caption{Dirty training, no pre-training.}
\label{fig:ea_dirty_none}
\end{subfigure}\\
\begin{subfigure}[b]{\myfigwidth}
\centering
\includegraphics[width=\textwidth]{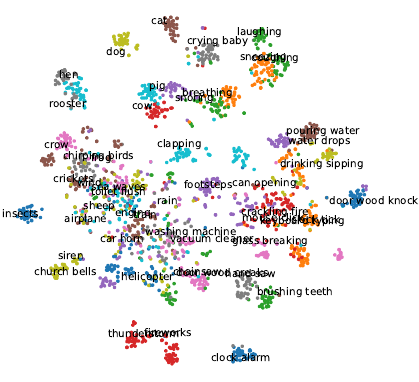}
\caption{Clean training, dirty pre-training.}
\label{fig:ea_clean_dirty}
\end{subfigure}
\begin{subfigure}[b]{\myfigwidth}
\centering
\includegraphics[width=\textwidth]{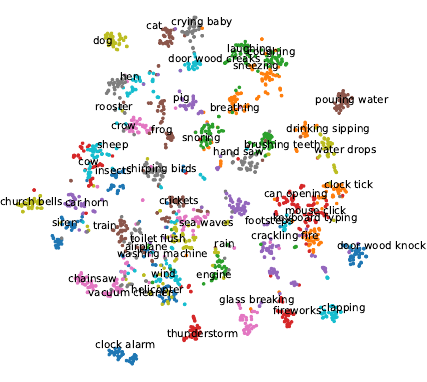}
\caption{Clean training, clean pre-training.}
\label{fig:ea_clean_clean}
\end{subfigure}
\begin{subfigure}[b]{\myfigwidth}
\centering
\includegraphics[width=\textwidth]{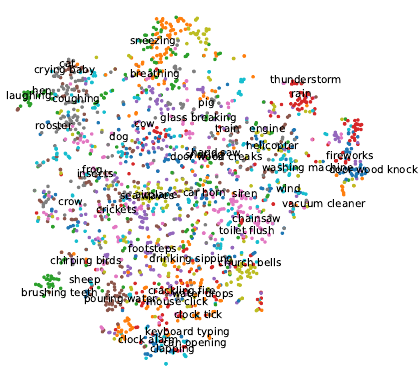}
\caption{Clean training, no pre-training.}
\label{fig:ea_clean_none}
\end{subfigure}
\caption{T-SNE projections of audio embeddings in the multimodal space ($E_a$) for each experiment. Data leakage from the dirty data results in better-defined clusters. When data leakage is eliminated, the clusters are not as well-resolved.}
\label{fig:audio_ea_tsne_projections}
\end{figure*}

\subsection{Zero-shot accuracy correlates with clusters in the audio embeddings}





The results in Table \ref{tab:results_esc50} indicate that ``dirty'' datasets lead to higher accuracy in zero-shot classification. This is expected, as it means that more information from the supposedly unknown labels is being leaked to the network at training time. 

The T-SNE projections \cite{tsne} corresponding to the embeddings $E_a$ for each experiment, shown in Figure~\ref{fig:audio_ea_tsne_projections}, reflect the same behavior. As it can be seen, ``dirty'' datasets seem to have better-defined clusters; conversely, the absence of pre-training leads to the most disorganized spaces. Importantly, pre-training with a clean dataset results in many well-defined clusters, which reflects the results shown in Table \ref{tab:results_esc50}.

Although the visual analysis can bring some insight into the data clusters, observations from these figures could be due to artifacts from the T-SNE projections. This issue was addressed by calculating the silhouette score for the $E_a$ in each experiment as shown in Table~\ref{tab:silhouette_esc50}. The silhouette score is calculated in the original dimension of $E_a$ (1024), hence it is not affected by distortions in the projection to a plane.

\begin{table}[!h]

\centering

\begin{tabular}{|c|c|c|c|} \hline

\multirow{2}{*}{Training} & 
\multicolumn{3}{|c|}{Pre-training} \\ \cline{2-4}

& 
Dirty & 
Clean & 
None \\ \hline

Dirty & 
$0.73$ & 
$0.61$ & 
$0.31$ \\

Clean & 
$0.48$ & 
$0.40$ & 
$0.14$ \\ \hline

\multicolumn{3}{|r|}{Random guess} & 
$0.02$ \\ \hline

\end{tabular}

\caption{Zero-shot accuracy results in the ESC50 dataset. As data gets more dirty, the system performance seems to increase because of data leakage.}

\label{tab:results_esc50}

\end{table}

\begin{table}[!h]

\centering

\begin{tabular}{|c|c|c|c|c|} \hline

\multicolumn{2}{|c|}{\multirow{2}{*}{Training}} & 
\multicolumn{3}{|c|}{Pre-training} \\ \cline{3-5}

\multicolumn{2}{|c|}{} &
Dirty &
Clean &
None \\ \hline

\multirow{2}{*}{Dirty} & 
$x_a$ &
$\phantom{-}0.35\phantom{-}$ &
$\phantom{-}0.20\phantom{-}$ &
$\phantom{-}0.11\phantom{-}$ \\

&
$E_a$ &
$\phantom{-}0.34\phantom{-}$ &
$\phantom{-}0.20\phantom{-}$ &
$\phantom{-}0.08\phantom{-}$ \\ \hline \hline

\multirow{2}{*}{Clean} &
$x_a$ &
$\phantom{-}0.27\phantom{-}$ &
$\phantom{-}0.16\phantom{-}$ &
$\phantom{-}0.01\phantom{-}$ \\

&
$E_a$ &
$\phantom{-}0.14\phantom{-}$ &
$\phantom{-}0.15\phantom{-}$ &
$-0.05\phantom{-}$ \\ \hline

\end{tabular}

\caption{Silhouette scores for audio encodings $x_a$ and embeddings $E_a$ in the cross-modal domain for each experiment in the ESC50 dataset using ground-truth labels as reference and cosine divergence as metric. The silhouette score is bounded between $-1$ and $1$, and higher scores are better.}

\label{tab:silhouette_esc50}

\end{table}

\begin{figure}[!h]
\centering
\includegraphics[width=0.27\textwidth]{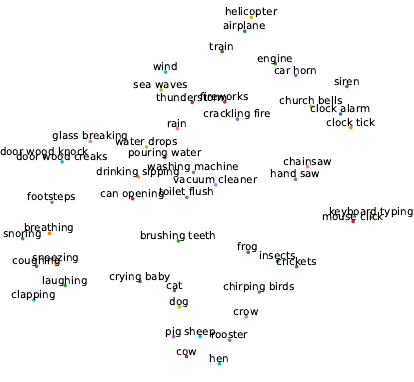}
\caption{T-SNE projection for ESC50 text embeddings in the cross-modal ($E_t$) space in the clean/clean experiment. As opposed to the audio embeddings shown in Figure~\ref{fig:audio_ea_tsne_projections}, captions are grouped according to the semantics of the labels, regardless of their sound similarity, indicating that audio and text embeddings in the cross-modal domain may carry different information.}
\label{fig:text_et}
\end{figure} 

\begin{figure*}[!t]
\centering
\begin{subfigure}[b]{\myfigwidth}
\includegraphics[width=\textwidth]{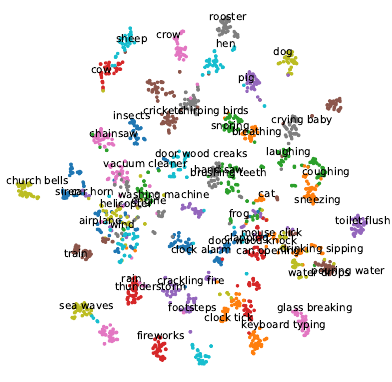}
\caption{Audio domain, dirty pre-training}
\label{fig:audio_tsne_projections_untrained_dirty}
\end{subfigure}
\begin{subfigure}[b]{\myfigwidth}
\includegraphics[width=\textwidth]{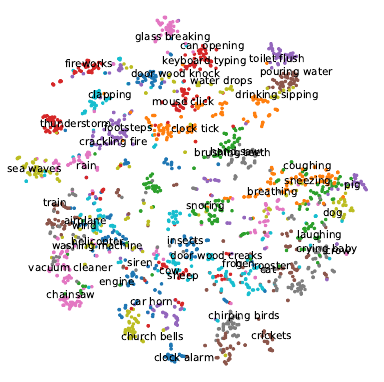}
\caption{Audio domain, clean pre-training}
\label{fig:audio_tsne_projections_untrained_clean}
\end{subfigure}
\begin{subfigure}[b]{\myfigwidth}
\includegraphics[width=\textwidth]{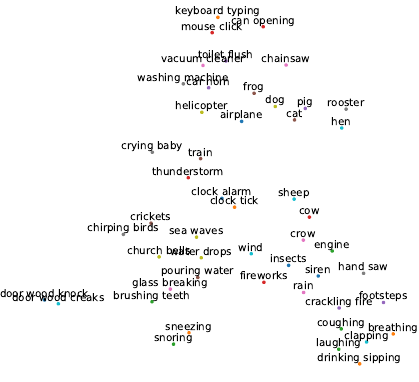}
\caption{Text domain}
\label{fig:text_tsne_projections_untrained}
\end{subfigure}
\caption{T-SNE projections for ESC50 embeddings $x_a$ and $x_t$ in the unimodal audio and text domain using the pre-trained backbones with no further fine tuning.}
\label{fig:original_audio}
\end{figure*}

These silhouette scores present a high Pearson correlation to the accuracy results shown in Table~\ref{tab:results_esc50} ($\rho=0.97, p<10^{-2}$). This correlation, together with the visual analysis of the T-SNE plots in Figure~\ref{fig:audio_ea_tsne_projections}, form a strong indication that zero-shot accuracy in this setting is essentially dependent on the presence of well-defined clusters in the audio embedding space. These clusters are further discussed next.


\subsection{The topology of text-domain embeddings}
One important aspect of $E_a$ for all pre-trained cases is that they span clusters that are related to the auditory aspect of each label. For example, Figure~\ref{fig:ea_clean_clean} has a cluster with items from the ``cow'' and ``sheep'' classes, whereas Figure~\ref{fig:ea_clean_dirty} has a cluster with ``crackling fire'' and ``mouse click''. Conversely, Figure~\ref{fig:text_et} shows that the text embeddings $E_t$ for the labels display a topology in which positions relate to the semantics of each label: ``door wood knock'', for example, is close to ``door wood creaks'', and ``clock alarm'' is close to ``clock tick'', even though, in both cases, the sounds of the recordings are clearly distinct.
Such similarities can be measured using a simple topology-structure-comparison algorithm, as follows. Let $P = \{p_i\}$ and $Q = \{q_i\}$, $i \in \{1, 2, \cdots, n\}$ be two point clouds to be compared, where points with the same index in both clouds correspond to the same entity (e.g. an audio-text pair). Define $N_{k}(p_i)$ to be the set of the indices of the $k$ closest points (including itself) to $p_i$ in $P$, and $N_{k}(q_i)$ analogously in $Q$. Then, the average value $S_{k}$ of $|N_{k}(p_i) \cap N_{k}(q_i)| / k$ over all indices is the topological similarity between $P$ and $Q$. The result $S_{k}$ is close to $1$ when all the neighbors are strictly the same, and close to $0$ when no neighbors are the same. Since $S_k$ changes with $k$, we use $S = \max_k S_k$ as the final topological similarity.

The topological-structure similarity between $x_a$ and $E_a$, $x_t$ and $E_t$, and $E_a$ and $E_t$ are respectively shown in Table~\ref{tab:topological_eaet}. To account for the difference in the number of samples in the audio domain (2000 samples) and the labels (50), we calculated the topological similarities using the label-wise mean vector of the audio representations.

\begin{table}[h]

\centering

\begin{tabular}{|c|c|c|c|c|c|} \hline

\multirow{2}{*}{Representations} & \multirow{2}{*}{Training}  & \multicolumn{3}{|c|}{Pre-training } \\ \cline{3-5}
& & Dirty & Clean & None \\ \hline
\multirow{2}{*}{$x_a$ vs $E_a$ }
& Dirty & $0.74$ & $0.75$ & $0.77$ \\
& Clean & $0.82$ & $0.78$ & $0.75$ \\
\hline  \hline
\multirow{2}{*}{$x_t$ vs $E_t$} 
& Dirty & $0.75$ & $0.72$ & $0.69$ \\
& Clean & $0.62$ & $0.75$ & $0.56$ \\
\hline \hline
 \multirow{2}{*}{$E_a$ vs $E_t$}
& Dirty & $0.46$ & $0.47$ & $0.44$ \\
& Clean & $0.50$ & $0.45$ & $0.38$ \\
\hline 
\end{tabular}
\caption{Topological structure similarity between item representations for each experiment. Values are bounded from $0$ to $1$, and higher numbers indicate a higher similarity. The similarity between the unimodal and its corresponding cross-modal representations ($x_a$ vs $E_a$ and $x_t$ vs $E_t$) is higher than the similarity between the cross-modal representations ($E_a$ vs $E_t$). This indicates that the cross-modal training was not able induce a common-ground domain from a topological perspective.}
\label{tab:topological_eaet}
\end{table}

Importantly, the topological similarity between $E_a$ and $E_t$ correlates with the logarithm of the zero-shot accuracy for each experiment (Pearson correlation, $\rho=0.85, p=0.032$). This means that, as the cross-modal space becomes a more suitable middle-ground representation for audio and text, there is a tendency for a higher accuracy in zero-shot classification.



\subsection{Cross-modal training does not resolve clusters}
As seen in Figure~\ref{fig:topology}, the contrastive cross-modal learning topology relies on two unimodal backbones. Each of these backbones is pre-trained in its own domain, and posteriorly fine-tuned jointly during the training of the cross-modal embeddings. Figure~\ref{fig:original_audio} shows $x_a$ and $x_t$ without the cross-modal training.

It is possible to see that the same clusters that appear unresolved in Figures \ref{fig:audio_tsne_projections_untrained_dirty} and \ref{fig:audio_tsne_projections_untrained_clean} still show as unresolved in the embeddings $E_a$ after training the cross-modal projections (Figures~\ref{fig:ea_dirty_dirty} and \ref{fig:ea_clean_dirty} for dirty pre-training, and Figures~~\ref{fig:ea_dirty_clean} and \ref{fig:ea_clean_clean} for clean pre-training). Remarkable examples are ``airplane'' and ``wind'' for the dirty pre-training, and ``cow'' and ``sheep'' for the clean pre-training. These observations are also supported by analyzing the silhouette scores for the $x_a$ projections in each experiment, shown in Table~\ref{tab:silhouette_esc50}. Remarkably, these silhouette scores are slightly higher than those related to $E_a$ (Table~\ref{tab:silhouette_esc50}), which further indicates that the projection into the cross-modal layer does not improve the cluster quality. 



This means that the model was not able to learn to differentiate new audio classes through the influence of the text backbone after cross-modal training. Rather, its ability to find clusters related to audio classes comes straight from the pre-training stages. Cross-modal training can slightly improve the cluster quality (from $x_a$ to $E_a$ spaces, see Table~\ref{tab:silhouette_esc50}), but not to the point of finding new, relevant decision borders.

Conversely, in the initial state of the text embeddings (Figure~\ref{fig:text_tsne_projections_untrained}), there are topological aspects that resemble the ones found after training and fine-tuning (Figure~\ref{fig:text_et}). It is possible to see that labels are grouped according to their semantics, that is, ``door wood knock'' is close to ``door wood creaks'' and ``clock alarm'' is close to ``clock tick''. This means that the cross-modal training was also unable to make significant topological changes in the text domain.

\section{Conclusion}
\label{sec:majhead}
The results discussed in this paper indicate that the zero-shot learning accuracy in the cross-modal network depends essentially on the clustering quality of the unimodal backbone embeddings. This means that a backbone pre-trained for a supervised classification task on the labels used in the downstream zero-shot classification task is prone to achieve outstanding accuracy. However, using this setup defeats the purpose of zero-shot learning itself, as the corresponding backbone could simply be used as a supervised classifier for the desired classes.

Through a case study, we have shown that T-SNE visualizations, silhouette scores, and an analysis of topological similarity can provide important information towards identifying the sources of higher accuracy in zero-shot classification in these cross-modal networks. These tools can also be used to identify data leakage from pre-trained networks without having access to the training data itself.

We hope that researchers in this space consider these ideas in future investigations in contrastive cross-modal training in order to further validate results usually reported as performance in benchmark tasks. This also suggests an opportunity to formulate better-informed standards for evaluating work in cross-modal zero-shot classification.

\pagebreak






\bibliographystyle{ieeebib}
\bibliography{refs}

\begin{thebibliography}{10}

\bibitem{pourpanah_review_2022}
Farhad Pourpanah, Moloud Abdar, Yuxuan Luo, Xinlei Zhou, Ran Wang, Chee~Peng
  Lim, Xi-Zhao Wang, and Q.~M.~Jonathan Wu,
\newblock ``A {Review} of {Generalized} {Zero}-{Shot} {Learning} {Methods},''
\newblock {\em IEEE Transactions on Pattern Analysis and Machine Intelligence},
  pp. 1--20, 2022.

\bibitem{lampert_learning_2009}
Christoph~H. Lampert, Hannes Nickisch, and Stefan Harmeling,
\newblock ``Learning to detect unseen object classes by between-class attribute
  transfer,''
\newblock in {\em 2009 {IEEE} {Conference} on {Computer} {Vision} and {Pattern}
  {Recognition}}, Miami, FL, June 2009, pp. 951--958, IEEE.

\bibitem{watanabe_query-by-blending_2019}
Kento Watanabe and Masataka Goto,
\newblock ``Query-by-{Blending}: {A} {Music} {Exploration} {System} {Blending}
  {Latent} {Vector} {Representations} of {Lyric} {Word}, {Song} {Audio}, and
  {Artist},''
\newblock in {\em International Society for Music Information Retrieval
  Conference}, 2019.

\bibitem{xian_zero-shot_2020}
Y.~Xian, B.~Schiele, and Z.~Akata,
\newblock ``Zero-shot learning — the good, the bad and the ugly,''
\newblock in {\em 2017 IEEE Conference on Computer Vision and Pattern
  Recognition (CVPR)}, Los Alamitos, CA, USA, jul 2017, pp. 3077--3086, IEEE
  Computer Society.

\bibitem{radford_clip_2021}
Alec Radford, Jong~Wook Kim, Chris Hallacy, Aditya Ramesh, Gabriel Goh,
  Sandhini Agarwal, Girish Sastry, Amanda Askell, Pamela Mishkin, Jack Clark,
  Gretchen Krueger, and Ilya Sutskever,
\newblock ``Learning transferable visual models from natural language
  supervision,''
\newblock {\em CoRR}, vol. abs/2103.00020, 2021.

\bibitem{elizalde2022clap}
Benjamin Elizalde, Soham Deshmukh, Mahmoud~Al Ismail, and Huaming Wang,
\newblock ``Clap learning audio concepts from natural language supervision,''
\newblock in {\em ICASSP 2023 - 2023 IEEE International Conference on
  Acoustics, Speech and Signal Processing (ICASSP)}, 2023, pp. 1--5.

\bibitem{xie_zero-shot_2019}
Huang Xie and Tuomas Virtanen,
\newblock ``Zero-{Shot} {Audio} {Classification} {Based} {On} {Class} {Label}
  {Embeddings},''
\newblock in {\em 2019 {IEEE} {Workshop} on {Applications} of {Signal}
  {Processing} to {Audio} and {Acoustics} ({WASPAA})}, New Paltz, NY, USA, Oct.
  2019, pp. 264--267, IEEE.

\bibitem{deshmukh2022audio}
Soham Deshmukh, Benjamin Elizalde, and Huaming Wang,
\newblock ``Audio retrieval with wavtext5k and clap training,''
\newblock in {\em INTERSPEECH 2023}, 2023.

\bibitem{Yusong2023-laionclap}
Yusong Wu, Ke~Chen, Tianyu Zhang, Yuchen Hui, Taylor Berg-Kirkpatrick, and
  Shlomo Dubnov,
\newblock ``Large-scale contrastive language-audio pretraining with feature
  fusion and keyword-to-caption augmentation,''
\newblock in {\em ICASSP 2023 - 2023 IEEE International Conference on
  Acoustics, Speech and Signal Processing (ICASSP)}, 2023, pp. 1--5.

\bibitem{choi_zero-shot_2019}
Jeong Choi, Jongpil Lee, Jiyoung Park, and Juhan Nam,
\newblock ``Zero-shot {Learning} for {Audio}-based {Music} {Classification} and
  {Tagging},''
\newblock in {\em International Society for Music Information Retrieval
  Conference}, 2019.

\bibitem{mercea_audiovisual_2022}
Otniel-Bogdan Mercea, Lukas Riesch, A.~Sophia Koepke, and Zeynep Akata,
\newblock ``Audiovisual {Generalised} {Zero}-shot {Learning} with {Cross}-modal
  {Attention} and {Language},''
\newblock in {\em 2022 {IEEE}/{CVF} {Conference} on {Computer} {Vision} and
  {Pattern} {Recognition} ({CVPR})}, New Orleans, LA, USA, June 2022, pp.
  10543--10553, IEEE.

\bibitem{wu2023}
Yusong Wu, Ke~Chen, Tianyu Zhang, Yuchen Hui, Taylor Berg-Kirkpatrick, and
  Shlomo Dubnov,
\newblock ``Large-scale contrastive language-audio pretraining with feature
  fusion and keyword-to-caption augmentation,''
\newblock in {\em ICASSP 2023 - 2023 IEEE International Conference on
  Acoustics, Speech and Signal Processing (ICASSP)}, 2023, pp. 1--5.

\bibitem{tsne}
Laurens van~der Maaten and Geoffrey Hinton,
\newblock ``Visualizing data using t-sne,''
\newblock {\em Journal of Machine Learning Research}, vol. 9, no. 86, pp.
  2579--2605, 2008.

\bibitem{Rousseeuw1987}
Peter~J. Rousseeuw,
\newblock ``Silhouettes: A graphical aid to the interpretation and validation
  of cluster analysis,''
\newblock {\em Journal of Computational and Applied Mathematics}, vol. 20, pp.
  53–65, Nov. 1987.

\bibitem{ramesh_hierarchical_2022}
Aditya Ramesh, Prafulla Dhariwal, Alex Nichol, Casey Chu, and Mark Chen,
\newblock ``Hierarchical {Text}-{Conditional} {Image} {Generation} with {CLIP}
  {Latents},''
\newblock 2022.

\bibitem{cnn14}
Qiuqiang Kong, Yin Cao, Turab Iqbal, Yuxuan Wang, Wenwu Wang, and Mark~D.
  Plumbley,
\newblock ``Panns: Large-scale pretrained audio neural networks for audio
  pattern recognition,''
\newblock {\em IEEE/ACM Transactions on Audio, Speech, and Language
  Processing}, vol. 28, pp. 2880--2894, 2020.

\bibitem{bert}
Jacob Devlin, Ming{-}Wei Chang, Kenton Lee, and Kristina Toutanova,
\newblock ``{BERT:} pre-training of deep bidirectional transformers for
  language understanding,''
\newblock in {\em Proc. of the 2019 Conference of the North American Chapter of
  the Association for Computational Linguistics 2019}. 2019, pp. 4171--4186,
  Association for Computational Linguistics.

\bibitem{clotho}
Konstantinos Drossos, Samuel Lipping, and Tuomas Virtanen,
\newblock ``Clotho: an audio captioning dataset,''
\newblock in {\em ICASSP 2020 - 2020 IEEE International Conference on
  Acoustics, Speech and Signal Processing (ICASSP)}, 2020, pp. 736--740.

\bibitem{audiocaps}
Chris~Dongjoo Kim, Byeongchang Kim, Hyunmin Lee, and Gunhee Kim,
\newblock ``Audiocaps: Generating captions for audios in the wild,''
\newblock in {\em NAACL-HLT}, 2019.

\bibitem{fonseca2022FSD50K}
Eduardo Fonseca, Xavier Favory, Jordi Pons, Frederic Font, and Xavier Serra,
\newblock ``{FSD50K}: an open dataset of human-labeled sound events,''
\newblock {\em IEEE/ACM Transactions on Audio, Speech, and Language
  Processing}, vol. 30, pp. 829--852, 2022.

\bibitem{macs}
Irene Mart{\'{\i}}n{-}Morat{\'{o}} and Annamaria Mesaros,
\newblock ``What is the ground truth? reliability of multi-annotator data for
  audio tagging,''
\newblock in {\em 29th European Signal Processing Conference, {EUSIPCO} 2021,
  Dublin, Ireland, August 23-27, 2021}. 2021, pp. 76--80, {IEEE}.

\bibitem{piczak2015dataset}
Karol~J. Piczak,
\newblock ``{ESC}: {Dataset} for {Environmental Sound Classification},''
\newblock in {\em Proceedings of the 23rd {Annual ACM Conference} on
  {Multimedia}}. pp. 1015--1018, {ACM Press}.

\bibitem{audioset}
Jort~F. Gemmeke, Daniel P.~W. Ellis, Dylan Freedman, Aren Jansen, Wade
  Lawrence, R.~Channing Moore, Manoj Plakal, and Marvin Ritter,
\newblock ``Audio set: An ontology and human-labeled dataset for audio
  events,''
\newblock in {\em 2017 IEEE International Conference on Acoustics, Speech and
  Signal Processing (ICASSP)}, 2017, pp. 776--780.

\bibitem{xie_zero-shot_2021}
Huang Xie and Tuomas Virtanen,
\newblock ``Zero-{Shot} {Audio} {Classification} {Via} {Semantic}
  {Embeddings},''
\newblock {\em IEEE/ACM Transactions on Audio, Speech, and Language
  Processing}, vol. 29, pp. 1233--1242, 2021.

\end{thebibliography}

\pagebreak

\end{document}